\documentclass[graybox]{svmult}

\usepackage{mathptmx}       
\usepackage{helvet}         
\usepackage{courier}        
\usepackage{type1cm}        
\usepackage{makeidx}         
\usepackage{graphicx}        
\usepackage{multicol}        

\makeindex             

\begin{document}

\title*{A Comprehensive Analysis of Time Series Segmentation on
the Japanese Stock Prices}
\titlerunning{A Comprehensive Analysis of Time Series Segmentation...}
\author{Aki-Hiro Sato}
\institute{Department of Applied Mathematics and Physics, Graduate
School of Informatics, Kyoto University,
\email{sato.akihiro.5m@kyoto-u.ac.jp}}
%
%
\maketitle

\abstract*{This study conducts a comprehensive analysis of time series
segmentation on the Japanese stock prices listed on the first section
of the Tokyo Stock Exchange during the period from 4 January 2000 to 30
January 2012. A recursive segmentation procedure is used under
the assumption of a Gaussian mixture. The daily number of each quintile of
volatilities for all the segments is investigated empirically. It
is found that from June 2004 to June 2007, a large majority of stocks are
stable and that from 2008 several stocks showed instability. On March
2011, the daily number of instable securities steeply increased due to
societal turmoil influenced by the East Japan Great Earthquake. It is
concluded that the number of stocks included in each quintile of
volatilities provides useful information on macroeconomic situations.}

\abstract{This study conducts a comprehensive analysis of time series
segmentation on the Japanese stock prices listed on the first section
of the Tokyo Stock Exchange during the period from 4 January 2000 to 30
January 2012. A recursive segmentation procedure is used under
the assumption of a Gaussian mixture. The daily number of each quintile of
volatilities for all the segments is investigated empirically. It
is found that from June 2004 to June 2007, a large majority of stocks are
stable and that from 2008 several stocks showed instability. On March
2011, the daily number of instable securities steeply increased due to
societal turmoil influenced by the East Japan Great Earthquake. It is
concluded that the number of stocks included in each quintile of
volatilities provides useful information on macroeconomic situations.}


\section{Introduction}
\label{sec:intro}
Recently, interests in a relationship between price
movements and events or news seem to
increase~\cite{Kenett:12,Ilaria,Preis}. The traded prices are assumed
to be signals resulting from both exogenous and endogenous factors.

Agent-based models of financial markets have been developed for the
last
decade~\cite{Zhang:98,Challet:00,Bornholdt:01,Krawiecki:02,Alfrano:2005,Feng}. Normally
buyers and sellers are assumed in the model. A market including
several agents is considered. Fundamentalists, chartists, and noise
traders may interplay in financial markets. The fundamentalists know actual
value of stock but the actual value fluctuates in time. The chartists
watch price movements and determine their prices based on the price
movements. The chartists are classified into trend followers and
contrarians. The noise traders determine their trading prices randomly.
As modelled above, the financial market consists of various types of
participants is a kind of agent-based information processing systems
embedded in the real world. 

In the last two decades, statistical properties of asset price returns 
have been successively studies in the literature of
econophysics~\cite{Mantegna, Bouchaud}. One of the important properties
is that the probability distribution of returns exhibits a fat-tailed
distribution~\cite{Mandelbrot,Gopikrishnan}. Several researchers
reported that the tail distributions of log returns have the
power-law and are well fitted by Student's
$t$-distributions~\cite{Praetz} or $q$-Gaussian 
distributions founded in nonextensive statistical thermo
dynamics~\cite{Tsallis,Aki-Hiro:10}. The Beck model introduced as a  
dynamical foundation of Tsallis statistics~\cite{Beck:Dyn}
Although it is originally intended to describe mechanical systems such
as turbulence, its basic idea of fluctuating temperature is well
consistent with the heteroscedasticity of markets. Thus, in recent
studies, it is employed to elucidate price fluctuations in
markets~\cite{Kozuki,Ausloos} and it is called
superstatistics~\cite{Beck:Sup}. Moreover, the boom-bust cycle is
associated with the existence of bubbles in stock markets. This cycle
may be quantified by collective behavior of stock prices. Researchers
have shown that there is some degree of collective behavior and
synchronization in the return of actual stocks~\cite{Shapira}.

Macroeconomics situations strongly influence money flows at all the
levels of society. Stocks at each sector are traded by investors and
traders through the stock exchange markets every minute. Moreover, stock
prices are so sensitive to the money flows that stock prices of all the
sectors depend on demand-supply situations of the money by economic
actors. Therefore, they are expected to be useful for detecting a
changes of macroeconomic situations. 

In this study, we hope to provide some insights on the problem on
quantification of Japanese macroeconomic situations through a
comprehensive analysis of stock prices traded in the Tokyo Stock
Exchange. In the context of economics and finance, there are various
methods to segment highly nonstationary financial time series into
stationary segments called regimes or trends. Following the pioneering
works by Goldfeld and Quandt~\cite{Goldfeld}, there is an enormous
literature on detecting structural breaks or change points separating
stationary segments. Recently, a recursive entropic scheme to segment
financial time series was proposed~\cite{Siew:11}. In fact they investigate
segments for Japanese stock indices, however, they did not consider
stock prices themselves. 

A recursive segmentation procedure is applied to analyze security
prices of 1,413 Japanese firms listed on the first section of the
Tokyo Stock Exchange. In this paper, the number of segments in
quintiles in terms of variance is computed in order to detect change
points of money flows of the Japanese security market.

This article is organized as follows. In Sec. \ref{sec:method}, the
recursive segmentation procedure is briefly explained. In
Sec. \ref{sec:numerical} the segmentation procedure is performed for
artificial time series. In Sec. \ref{sec:empirical}, the empirical
analysis with daily log-returns for the last 10 years is
conducted. Sec. \ref{sec:conclusion} is devoted to conclusion.

\section{Method}
\label{sec:method}
\subsection{Segmentation procedure}

Let $O_{i,t}$ and $E_{i,t}$ be, respectively, daily opening and
closing prices of $i$-th stock $(i=1,\ldots,M)$ at day $t \quad
(t=1,\ldots,n)$. $M$ and $n$ are denoted as the total number of stock
and the total number of observations. The daily log-return (opening to
closing) time series $x_{i,t} \quad (i=1,\ldots,M; t=1,\ldots,n)$ is
computed as
\begin{equation}
x_{i,t} = \log E_{i,t} - \log O_{i,t}.
\label{eq:log-ret}
\end{equation}

According to the seminal work by Mantegna and Stanley~\cite{Mantegna},
the log-return time series of stock prices are modeled by L\'evy
distributions. Superstatistics suggests that a mixture of Gaussian
distributions with $\chi^2$-distributions in terms of variance gives
a L\'evy distribution. Therefore, we may assume that each segment is
sampled from a Gaussian distribution with different mean and
variance. Namely, it is assumed that the log-return time series consist
of $m_i$ stationary segments. Each segment follows a stationary Gaussian
distribution with mean $\mu_{i,j}$ and variance $\sigma_{i,j}^2$
($j=1,\ldots,m_i$).

To find the $m_i-1$ unknown segment boundaries $t_{i,j}$ separating
segment $j$ and $j+1$, the recursive segmentation scheme introduced
by Siew et al and Bernaola et al~\cite{Siew:11,Bernaola}. Their
segmentation scheme is fundamentally based on the likelihood-ratio
test under an {\it i.i.d} Gaussian null model and a joint consisting
two different Gaussian models for the total time series.

Firstly, suppose that there are $n$ observations $x_s \quad (s=1,\ldots,n)$. 
Let $g(x;\mu,\sigma^2)$ be a Gaussian distribution
\begin{equation}
g(x;\mu,\sigma^2) =
 \frac{1}{\sqrt{2\pi\sigma^2}}\exp\Bigl[-\frac{(x-\mu)^2}{2\sigma^2}\Bigr],
\label{eq:gauss}
\end{equation}
with parameters $\mu$ and $\sigma^2$. Assuming that the
observations $x_s$ should be segmented at $t$ and that the observations
on the left hand side are sampled from $g(x;\mu_L,\sigma_L^2)$, and 
ones on the right hand side are from $g(x;\mu_R,\sigma_R^2)$, we define
likelihood functions 
\begin{eqnarray}
L_1 &=&
 \prod_{s=1}^{n}g(x_s;\mu,\sigma^2), \\
\nonumber
L_2(t) &=& 
 \prod_{s=1}^{t}g(x_s;\mu_L,\sigma_L^2)\prod_{s=t+1}^{n}g(x_s;\mu_R,\sigma_R^2). \\
\end{eqnarray}
Furthermore, we define the logarithmic likelihood-ratio between $L_1$
and $L_2(t)$ as, 
\begin{equation}
\Delta(t) = \log L_2(t) - \log L_1\Bigr.
\label{eq:def-delta}
\end{equation}
Inserting Eq. (\ref{eq:gauss}) into Eq. (\ref{eq:def-delta}) we have
\begin{equation}
\Delta(t) = \sum_{s=1}^t\log
 g(x_s;\mu_L,\sigma^2_L) + \sum_{s=t+1}^n\log g(x_s;\mu_R,\sigma^2_R) - 
\sum_{s=1}^n\log g(x_s;\mu,\sigma^2).
\end{equation}
By using the approximations
\begin{eqnarray}
\nonumber
\frac{1}{n}\sum_{s=1}^n \log g(x_s;\mu,\sigma^2) &\approx&
 \int_{-\infty}^{\infty} g(x;\mu,\sigma^2) \log g(x;\mu,\sigma^2)
 \mbox{d}x = -\frac{1}{2}\log \Bigl(2\pi e \sigma^2\Bigr), \\
\nonumber
\frac{1}{t}\sum_{s=1}^t \log g(x_s;\mu_L,\sigma_L^2) &\approx&
 \int_{-\infty}^{\infty} g(x;\mu_L,\sigma_L^2) \log g(x;\mu_L,\sigma_L^2)
 \mbox{d}x = -\frac{1}{2}\log \Bigl(2\pi e \sigma_L^2\Bigr), \\
\nonumber
\frac{1}{n-t}\sum_{s=t+1}^{n} \log g(x_s;\mu_R,\sigma_R^2) &\approx&
 \int_{-\infty}^{\infty} g(x;\mu_R,\sigma_R^2) \log g(x;\mu_R,\sigma_R^2)
 \mbox{d}x = -\frac{1}{2}\log \Bigl(2\pi e \sigma_R^2\Bigr),
\end{eqnarray}
$\Delta(t)$ is rewritten as
\begin{equation}
\Delta(t) = n \log \sigma - t \log \sigma_L - (n-t)\log \sigma_R \geq 0.
\label{eq:delta1}
\end{equation}
$\sigma$, $\sigma_L$ and $\sigma_R$ are further approximated as maximum
likelihood estimators given by empirical standard deviations
\begin{eqnarray}
\sigma &=& \sqrt{\frac{1}{n}\sum_{s=1}^n x^2_s - \Bigl(\frac{1}{n}\sum_{s=1}^n
 x_s \Bigr)^2}, \\
\sigma_L &=& \sqrt{\frac{1}{t}\sum_{s=1}^t x^2_s - \Bigl(\frac{1}{t}\sum_{s=1}^t
 x_s \Bigr)^2}, \\
\sigma_R &=& \sqrt{\frac{1}{n-t}\sum_{s=t+1}^{n} x^2_s - \Bigl(\frac{1}{n-t}\sum_{s=t+1}^{n}
 x_s \Bigr)^2}.
\end{eqnarray}

$\Delta(t)$ can be used as an indicator to separate the observations
into two parts. An adequate way to separate the observations is
that segmentation is conducted at $t$ where $\Delta(t)$ takes the
maximum value. Namely, an adequate segmentation should be done at
\begin{equation}
t^* = \arg \max \Delta(t).
\end{equation}
If $\max \Delta(t)$ is less than a threshold value $\Delta_c$, then the
segmentation should be terminated. The hierarchical segmentation
procedure is also applied to the time series. After segmentation,
we also apply this procedure for each segment recursively. In order to 
stop the segmentation procedure we assume that the minimum value
$\Delta_c$. If $\max \Delta(t)$ is less than $\Delta_c$, then we do not
apply the segmentation procedure any more. This is used as the stopping
condition of the recursive segmentation procedure. Statistical theory
tells us that as the sample size $t$ approaches $\infty$, $2\Delta(t)$
is asymptotically $\chi^2$ distributed with degrees of freedom equal
to the difference between the number of alternative model parameters
and the number of null model parameters. In this Gaussian case, the
degrees of freedom is set as $2 = 4-2$. The cumulative distribution of
$\chi^2(x,2)$ is given by
\begin{equation}
F(x) = 1 - \exp\bigl(-\frac{x}{2}\bigr).
\end{equation}
Therefore, for the level of significance $\alpha (0 < \alpha < 1)$,
the threshold is given as
\begin{equation}
\Delta_c = -2\log (1-\alpha).
\end{equation}

Consequently, for stock $i \quad (i=1,\ldots,M)$, we obtain $m_i$
segments and $m_i$ sets of Gaussian parameters $\{\mu_{i,j}, \sigma_{i,j}^2\}
\quad (j=1, \ldots, m_i)$.

\subsection{General case}
More generally, let us define the discriminator $\Delta(t)$ given in
Eq. (\ref{eq:delta1}). Assume that $p(x;\theta)$ is a probability density 
function (model) parameterized by $\theta$. Let $\hat{\theta}$, 
$\hat{\theta}_L$, and $\hat{\theta}_R$ be maximum likelihood functions
computed by
\begin{eqnarray}
\hat{\theta} &=& \arg \max_{\theta} \sum_{s=1}^n \log p(x_s;\theta), \\
\hat{\theta}_L &=& \arg \max_{\theta} \sum_{s=1}^t \log p(x_s;\theta), \\
\hat{\theta}_R &=& \arg \max_{\theta} \sum_{s=t+1}^n \log p(x_s;\theta).
\end{eqnarray}
where $\hat{\theta}_L$ and $\hat{\theta}_R$ are maximum likelihood
estimators in left and right sequences, respectively. Assuming
likelihood functions
\begin{eqnarray}
L_1 &=& \prod_{s=1}^n p(x_s;\theta), \\
L_2(t) &=& \prod_{s=1}^t p(x_s;\theta_L)
 \prod_{s=t+1}^np(x_s;\theta_R),
\end{eqnarray}
we can rewrite Eq. (\ref{eq:def-delta}) as
\begin{eqnarray}
\nonumber
\Delta(t) &=& -n \int_{-\infty}^{\infty} p(x;\hat{\theta}) \log
p(x;\hat{\theta}) dx \\
&+& t \int_{-\infty}^{\infty} p(x;\hat{\theta}_L) \log p(x;\hat{\theta}_L) dx + (n-t) \int_{-\infty}^{\infty} p(x;\hat{\theta}_R) \log p(x;\hat{\theta}_R) dx,
\end{eqnarray}
$\Delta(t)/n$ is equivalent to
\begin{equation}
\frac{\Delta(t)}{n} = H\Bigl[p(x;\hat{\theta})\Bigr] - 
\frac{t}{n}H\Bigl[p(x;\hat{\theta}_L)\Bigr] -
\frac{n-t}{n}H\Bigl[p(x;\hat{\theta}_R)\Bigr],
\label{eq:JS}
\end{equation}
where $H[p(x;\theta)]$ represents the notation of Shannon entropy defined as
\begin{equation}
H\Bigl[p(x)\Bigr] = -\int_{-\infty}^{\infty} p(x;\theta) \log p(x;\theta) dx.
\end{equation}

In this Gaussian case, the degrees of freedom $k$ is set as the
dimensionality of a set of model parameters $\theta$. The cumulative 
distribution of $\chi^2$ distribution with degrees of freedom $k$ is
given as the regularized incomplete gamma function,
\begin{equation}
F(x) = \gamma\Bigl(\frac{k}{2},\frac{x}{2}\Bigr) =
\frac{1}{\Gamma\bigl(\frac{k}{2}\bigr)}
\int_{0}^{\frac{x}{2}}t^{\frac{k}{2}-1}\exp(-t)dt.
\end{equation}
The threshold for the terminated condition under a given level of
significance $\alpha$ is calculated from the regularized incomplete gamma
function, such that
\begin{equation}
\gamma\Bigl(\frac{k}{2},\frac{\Delta_c}{2}\Bigr) = \alpha.
\end{equation}

\section{Numerical study}
\label{sec:numerical}

Artificial time series are generated from {\it i.i.d} standard
normal random variables with different variances. Time series consisting
of 4 segments with different variance are generated. The time series in
the first segment is sampled from the normal distribution with mean 0
and standard deviation 1. The second is generated from the normal
distribution with mean 0 and standard deviation 2. The third is from
mean 0 and standard deviation 1. The fourth is from mean 0 and standard
deviation 3. The length of each segment is set as 500. Let $x(t)$ be
$T=2,000$ independent normal random variables with different variances
generated by means of the Box-Muller algorithm. $\Delta_c$ is fixed as
10 ($\alpha=0.99995$).

\begin{figure}[h]
\centering
\includegraphics[scale=0.9]{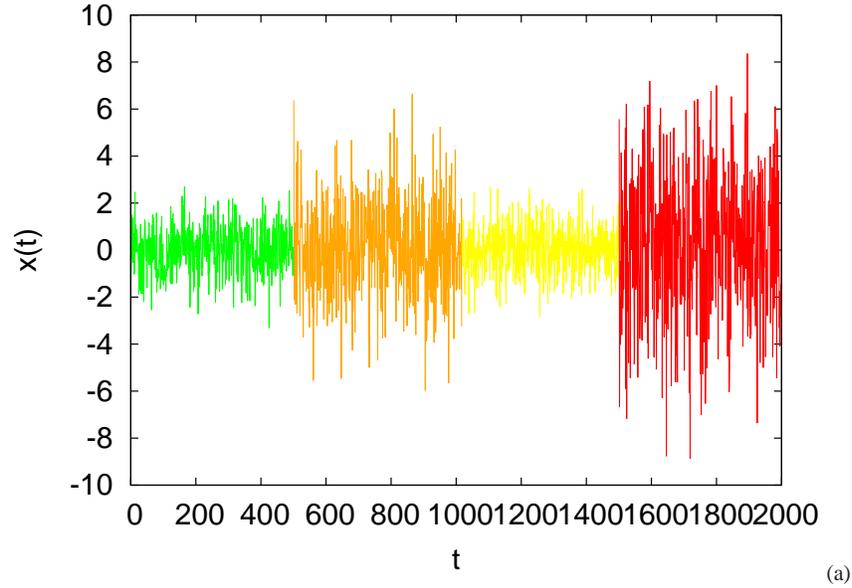}(a)
\caption{The time series consisting of 4 segments. Sequences in each
segment are sampled from a zero-mean Gaussian distribution. The standard
 deviation is set as 1, 2, 1, and 3 from the left segment. The length of
 each segment is set as 500.}
\label{fig:artificial}
\end{figure}

Fig. \ref{fig:artificial} shows the time series segmented by the
proposed hierarchical procedure. The length, mean and standard deviation
of each segment is  shown in Tab. \ref{tab:artificial}. From the table,
it is found that both the first and fourth segments are completely
determined. Both the second and third segments are slightly different
from the actual setting. The segments determined by the recursive
segmentation procedure are approximately ones assumed in advance. We
confirmed that the proposed procedure can determine the position where
each variance changed.

\begin{table}[h]
\centering
\caption{The length, mean and standard deviation of each segment
 detected by the proposed method.}
\label{tab:artificial}
\begin{tabular}{crrrrr}
\hline
segment & start & end & length & mean & std. dev. \\
\hline
1 & 1 & 500 & 500 & -0.00691 & 1.000 \\
2 & 501 & 1017 & 517 & 0.0524 & 1.986 \\
3 & 1018 & 1500 & 493 & 0.0320 & 0.980 \\
4 & 1501 & 2000 & 500 & 0.212 & 2.929 \\
\hline
\end{tabular}
\end{table}

\section{Empirical analysis}
\label{sec:empirical}

1,413 companies listed on the first section of Tokyo Stock
Exchange are selected for empirical analysis ($M=1,413$). The duration
is 4th January 2000 to 30 January 2012. There are 2,675 business days during the
observation period ($n=2,675$). These companies last during the observation
period. I apply the recursive segmentation procedure to 1,413 security
prices. I conducted the segmentation analysis of daily log-return time
series between daily opening and ending prices defined in
Eq. (\ref{eq:log-ret}). Throughout the investigation $\Delta_c$ is fixed as
10 ($\alpha = 0.99995$).

\begin{figure}[h]
\centering
\includegraphics[scale=0.9]{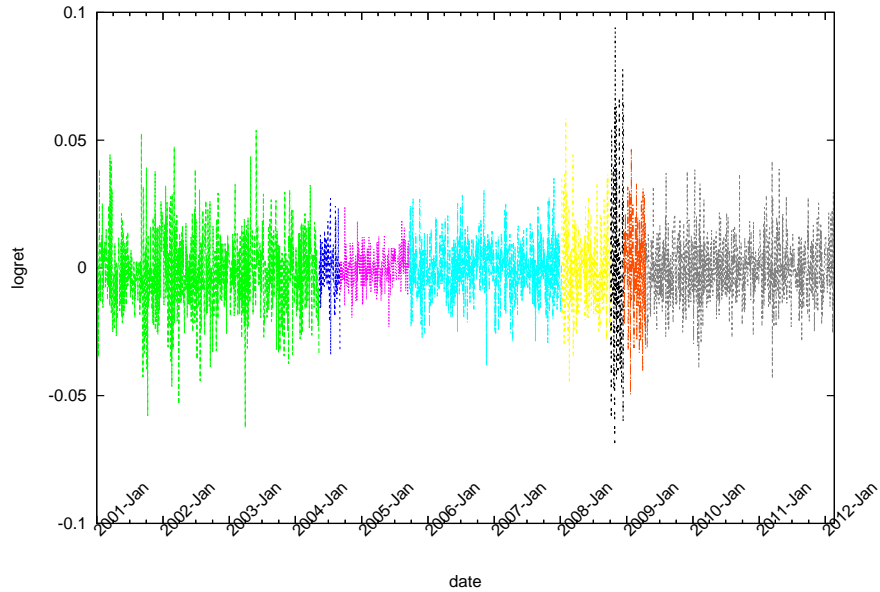}
\caption{The daily log return time series of Toyota Motor Corp (7203)
during the period from 4th January 2000 to 30th January 2012.}
\label{fig:Toyota}
\end{figure}

Fig. \ref{fig:Toyota} shows the daily log return time series of Toyota
Motor Corp (7203) segmented by using the proposed procedure. 9
segments are obtained in this case. The boundaries are computed
in 2000-07-03, 2004-05-10, 2004-09-03, 2005-09-16, 2008-01-08,
2008-10-03, 2008-12-15, and 2009-04-16.

\begin{figure}[h]
\centering
\includegraphics[scale=0.9]{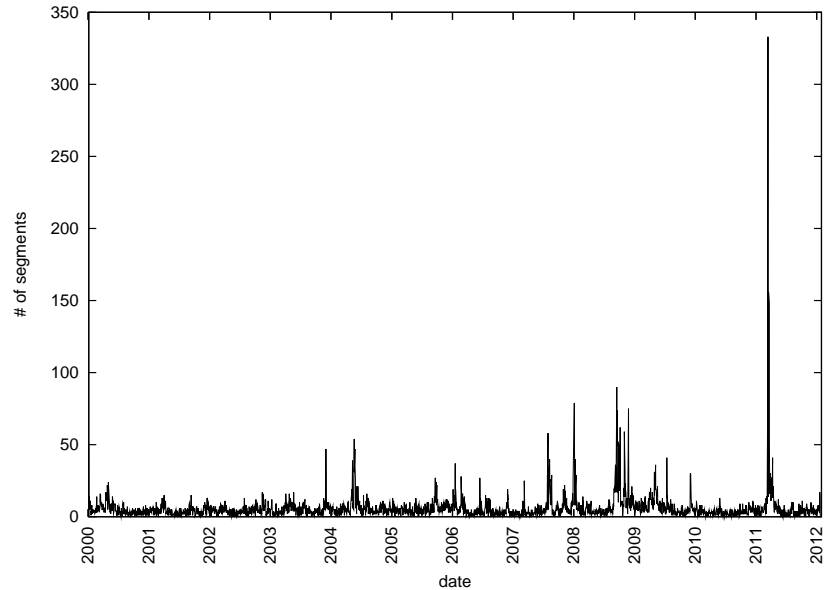}
\caption{The number of staring dates of segments for 1,413 log-return
 time series during the period from 4th January 2000 to 30th January 2012.}
\label{fig:count}
\end{figure}

Fig. \ref{fig:count} shows the number of starting dates of segments 
for 1,413 log-return time series. The number of segments increase
at (a) June 2000, (b) April 2004, (c) February 2006, and (d) 2007 to
2009. These seem to correspond to regimes or change points on Japanese
economy. Specifically, during the latest global financial crisis the
number of segments tend to increase (about 260 segments can be found at
this period). Furthermore, after (e) 11 March 2011, the Great East Japan
Earthquake, the number of segments steeply increased larger than the
latest global financial crisis.

\begin{figure}[h]
\centering
\includegraphics[scale=0.9]{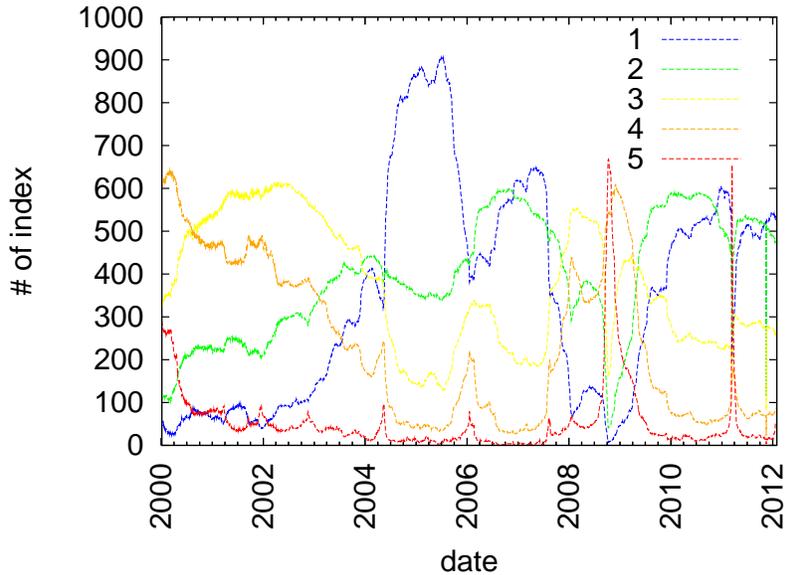}
\caption{The number of segments in each quintiles for 1,413 log-return
 time series during 4th January 2000 to 30th January 2012.}
\label{fig:quintile2}
\end{figure}

The number of segments belonging to the same quintile of variances is
counted. I computed order statistics of variance $\{\sigma_{i,(1)} \leq
\ldots \leq \sigma_{i,(m_i)}\}$ in $m_i$ segments of stock $i$. 
Next, each segment of stock $i$ is labeled $k=\{1,2,3,4,5\}$
depending on variance which belongs to the quintile. The number of
segments which have the same labels is counted at each
day. Fig. \ref{fig:quintile2} shows the number of segments belonging to
each quintile at every day. The number of first quintile segments 
shows stability of economic affairs, and the number of fifth quintile
segments indicates instability of economic affairs. It is found that
from 2003 to 2007 (I) the Japanese economy was in stable regime. From the
end of 2007 (II) the unstable regime was observed. Specifically September
2008 (III), when we experienced the Lehman shock, the number of fifth quintile
regimes steeply increased. This implies that the money flows of Japanese
economy became unstable just after the Lehman shock. From March 2009
(IV) the money flow eventually recovered and the number of unstable regimes
decreased and the number of stable segments eventually increased. From
11 March to 10 April 2011 (V), the number of unstable regimes steeply
increased due to the Great East Japan Earthquake. However, after that
the number of stable regimes rapidly increased and the Japanese
macroeconomic affairs have been recovered eventually. 

\begin{figure}[h]
\centering
\includegraphics[scale=0.9]{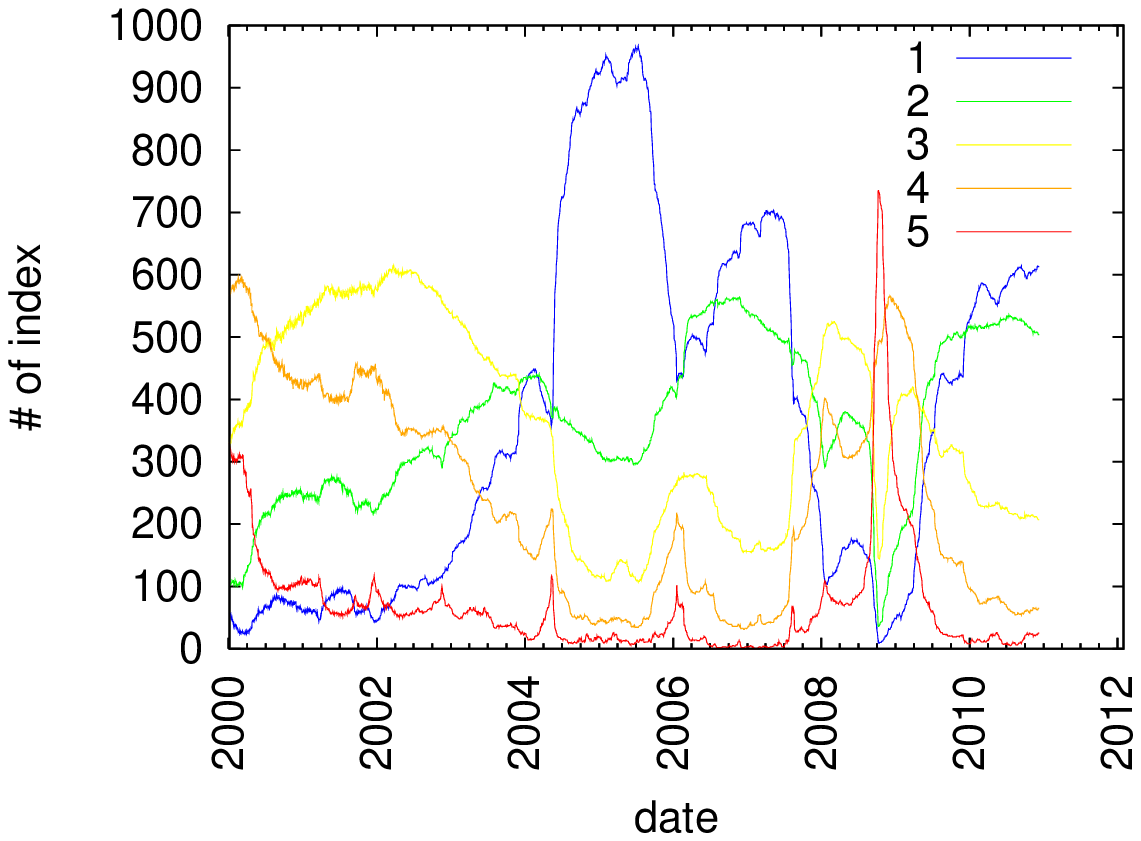}
\caption{The number of segments in each quintiles for 1,413 log-return
 time series during 4th January 2000 to 14th December 2010.}
\label{fig:quintile}
\end{figure}

Normally, we use only the past time series and want to foresee future
macroeconomic conditions. To do so, at least the obtained results should
be robust. In order to confirm the robustness, stability of the number of
each regime, we consider two different time periods. We use two time
periods for 2000-2010 (estimation) and for 2000-2012 (realization) and 
compare their results. The second is the same time series as ones shown in
Fig. \ref{fig:quintile2}. Fig. \ref{fig:quintile} shows the number of
regimes belonging to each quintile for two different periods. As seen
from them it is confirmed that they are almost same until 2010. Namely,
the time series may be employed to foresee future macroeconomic conditions.

\section{Conclusion}
\label{sec:conclusion}

A comprehensive time series segmentation analysis of the Japanese
security prices was conducted. The daily log-returns for 1,413 security
prices listed on the first section of the Tokyo Stock Exchange for 4
January 2000 to 14 December 2010 were analyzed by using a recursive
segmentation procedure based on Jensen-Shannon divergence. 

It was found that the number of segments increase at (a) June 2000, (b)
April 2004, (c) February 2006, (d) 2007 to 2009, and (e) March
2011. These seemed to correspond to regimes or change points on Japanese
economy.

The number of segments belonging to the same quintile of variances was
counted. It was found that from 2003 to 2007 (I) the Japanese economy
was in stable regime. From the end of 2007 (II) the unstable regime was
observed. Specifically September 2008 (III), when we experienced the Lehman
shock, the number of fifth quintile regimes steeply increased. This
implies that the money flows of Japanese economy became unstable just
after the Lehman shock. From March 2009 (IV) the money flow eventually
recovered and the number of unstable regimes decreased and the number
of stable segments eventually increased. After 11 March 2011 (V), the Great
East Japan Earthquake, the number of segments steeply increased. From 11
March to 10 April 2011, the macroeconomic situations seemed to be
negative, however, they were eventually recovered after that.

The volatility distribution computed from daily log-returns for a broad
spectrum of stock prices traded in the Tokyo Stock Exchange market
provides us with information on money flows at several economic levels. It
may be further useful to understand the macroeconomic affairs of Japan
in the quantitative way.

\end{document}